\documentclass{article}
\usepackage{amsmath,graphicx,color,booktabs,multirow,pifont}
\usepackage[T1]{fontenc} 
\usepackage[utf8]{inputenc} 
\usepackage{cite,url}
\usepackage{xcolor}
\usepackage{enumitem}
\usepackage{amsfonts}
\usepackage{dsfont} 
\usepackage{comment}
\usepackage{makecell} 
\usepackage{nicefrac}
\usepackage[bookmarks=false,hidelinks]{hyperref}
\usepackage{lineno}
\usepackage[preprint]{spconf}

\def\mycopyrightnotice{%
  \begin{minipage}{\textwidth}
 \scriptsize
  \copyright~2023 IEEE. Personal use of this material is permitted. Permission from IEEE must be obtained for all other uses, in any current or future media, including reprinting/republishing this material for advertising or promotional purposes, creating new collective works, for resale or redistribution to servers or lists, or reuse of any copyrighted component of this work in other works.
  \end{minipage}
}

\title{Similar but faster: Manipulation of tempo in music audio embeddings for tempo prediction and search}

\name{Matthew C. McCallum\textsuperscript{*}\thanks{\textsuperscript{*}equal contribution}, Florian Henkel\textsuperscript{*}, Jaehun Kim, Samuel E. Sandberg, Matthew E. P. Davies}
\address{SiriusXM-Pandora, USA}
\begin{document}
\ninept
\maketitle
\begin{abstract}

Audio embeddings enable large scale comparisons of the similarity of audio files for applications such as search and recommendation. Due to the subjectivity of audio similarity, it can be desirable to design systems that answer not only whether audio is similar, but similar in what way (e.g., wrt. tempo, mood or genre). Previous works have proposed disentangled embedding spaces where subspaces representing specific, yet possibly correlated, attributes can be weighted to emphasize those attributes in downstream tasks. However, no research has been conducted into the independence of these subspaces, nor their manipulation, in order to retrieve tracks that are similar but different in a specific way. Here, we explore the manipulation of tempo in embedding spaces as a case-study towards this goal. We propose tempo translation functions that allow for efficient manipulation of tempo within a pre-existing embedding space whilst maintaining other properties such as genre. As this translation is specific to tempo it enables retrieval of tracks that are similar but have specifically different tempi. We show that such a function can be used as an efficient data augmentation strategy for both training of downstream tempo predictors, and improved nearest neighbor retrieval of properties largely independent of tempo.

\end{abstract}

\copyrightnotice{\mycopyrightnotice}

\section{Introduction}
\label{sec:intro}

Generic audio embeddings enable search and organization of large scale audio catalogs with the benefit of computational efficiency as each audio waveform is analyzed only once, yet may be employed for any number of seen or unseen downstream tasks such as 
recommendation \cite{van2013deep}, labelling / categorization \cite{mccallum22ismir, DBLP:conf/ismir/HamelDYG13}, or audio similarity \cite{DBLP:conf/ismir/HamelDYG13}.  Recent advancements have demonstrated that such embeddings are not only general to a range of downstream tasks, but can achieve state-of-the-art performance in many of them \cite{mccallum22ismir, li2023mert, DBLP:conf/ismir/SpijkervetB21}. 

Tasks such as search, recommendation and similarity can often be highly contextual, subjective and personalized \cite{DBLP:conf/ismir/UrbanoDMS12, DBLP:journals/jiis/SchedlFU13}. As such it is desirable to have interpretable embeddings that can be manipulated or emphasized with respect to certain audio characteristics. In this way a user may be able to specify not only that they are interested in similar audio, but similar with emphasis on specific attributes. Alternatively, users may look for audio that is distinctly different in certain attributes but similar in others. For example, professional music producers and DJs may be looking for audio that has a similar feel to a song they are familiar with, but at a different, specific tempo. Other users may wish to explore properties such as instrumentation, key, genre etc. depending on their needs.

Previous research into disentangled embedding spaces have shown promising results in this direction \cite{DBLP:conf/ismir/LeeBSJN20, DBLP:conf/sigir/XunZYZDZDLZ023, lee2020disentangled}, where audio embeddings have been designed by training subspaces of the embedding space on specific characteristics such as genre, mood, instruments and tempo. In \cite{lee2020disentangled}, it is shown that predictions based on these individual subspaces can achieve excellent performance in the task relating to its characteristics. It is proposed that embeddings designed in this way could be weighted differently wrt. each subspace such that certain signal characteristics are emphasized in downstream applications such as nearest neighbor search. In applications where attribute weighted search is important, it is interesting not only to study the predictive performance of each subspace for their respective intended attribute, but also the independence of each subspace to alternative attributes. For example, it may be desirable to have a genre subspace that is not only highly predictive of genre, but impartial to other attributes such as tempo and instrumentation, unless those attributes are highly correlated with genre. The ability to design embeddings that can be manipulated to be attentive to certain attributes but independent of others, is important for search and discovery as it allows downstream users to direct the intention of their search to certain properties of interest and explore differences in others.

In this work, we adopt a different, novel approach to exploring interpretable embedding properties. From a high-level perspective, we take a state-of-the-art audio embedding that performs well on a task pertaining to a specific property, and design a translation function in the embedding space to manipulate that property, whilst maintaining other uncorrelated or loosely correlated properties. In this sense, the translation function describes manipulations of that property as contours in the embedding space. This allows a pre-existing embedding to be manipulated directly, providing access to embeddings describing not only the audio from which it was constructed but variations of it wrt. a certain property. 

While it is possible to conceive such translation functions for a variety of musical traits, in this paper we consider musical tempo as our dimension of interest, which is shown to be particularly salient in \cite{mccallum24icassp}. We show how its direct manipulation in the embedding space can enable the following downstream applications:

\textbf{(1) Nearest neighbor retrieval of a specific tempo}. Using the translation function to change only the tempo of an embedding, while maintaining other properties, allows us to search for audio that has similar musical characteristics, but a different tempo (Section~\ref{subsec:nnr_tempo}). 

\textbf{(2) Nearest neighbor retrieval impartial to tempo.} By not only employing nearest neighbor search of a single embedding, but along the \emph{embedding contour} with respect to tempo, we improve nearest neighbor retrieval of audio with properties that are largely independent of tempo (Section~\ref{subsec:nnr_impartial}).

\textbf{(3) Data augmentation for downstream tempo labelling.} 
Compared to directly augmenting the audio or spectrogram input during training of  a tempo prediction model, the translation function can be used as an efficient data augmentation strategy to improve downstream tempo labelling performance (Section \ref{subsec:data_aug}).

Finally, we hypothesize that the translation function can be achieved with lower computational complexity than the embedding model itself avoiding costly retraining of embedding models, or re-analysis of manipulated audio, making it a more pragmatic approach for exploring pre-existing embedding spaces.

\section{Methodology}

\subsection{Music Audio Embeddings}

In this work we employ the open-source MULE model \cite{mccallum22ismir} which achieves state-of-the-art performance on a variety of MIR tasks. This model is trained via contrastive learning \cite{chen2020simclr} using randomly sampled pairs of mel-spectrogram excerpts that are no more than $10$ seconds apart within the same track.
MULE encodes $3$\,s of audio into $1728$-dimensional embeddings which can either be used on their own (excerpt-level embeddings), or as a single embedding for a track by averaging across the time-line (track-level embeddings), without negatively impacting the performance on potential downstream tasks.
In the following, we will utilize both variants by using excerpt-level embeddings when training the translation function and track-level embeddings for nearest neighbor retrieval.

The input to this model are mel-spectrogram excerpts:
\begin{equation}
    \resizebox{0.90\hsize}{!}{
$X[u,m] = \log_{10}\left(\sum\limits_{k=0}^{k=K} S_u[k] \left\vert \sum\limits_{n=0}^{n=N}{e^{-\frac{2 \pi n k j}{K}}w[n]x[ml-n]} \right\vert \right)$ ,
}
\end{equation}
with $S_u[k]$ being a mel window at index $u$ for $0\leq u < U$ following the HTK mel scaling \cite{young2002htk}, $K$ being the DFT size, and $l$ the hop size. $w[n]$ denotes a Hann window of size $N$. For the computation of the mel-spectrograms we use the same parameters as described in \cite{mccallum22ismir}.

Subsequently, we refer to an embedding of these excerpts as 
\begin{equation}
z  = g(X[u,m]),
\end{equation}
with $g$ denoting the embedding network.

\subsection{Learning a Tempo Translation Function}
The goal of a tempo translation function is to directly manipulate an embedding $z$ such that the \emph{translated} embedding $z'$ maintains the same properties as the initial embedding while only changing the original (unknown) tempo $T$ by a stretch factor $\tau$, i.e., $T'=\tau T$.

Learning such a function can be done entirely self-supervised by artificially creating translated embeddings $z'$ as training targets. To this end we sample random mel-spectrogram excerpts $X[u,m]$ and perform time stretching via resampling of $X[u,m]$ using cubic spline interpolation at points $t = \tau m$ similar to \cite{schreiber18ismir}, i.e.,
\begin{equation}
    X_{TS}[u,m]=h_{TS}(X[u,m]; \tau),
\end{equation}
with stretching function, $h_{TS}$.
The time stretch factor $\tau$ is sampled in the range $[0.75,1.5]$ according to a log-uniform distribution
\begin{equation}
    \mathbf{\tau} \sim \frac{1}{\tau \log{(\tau_{max}/\tau_{min})}},
\end{equation}
with $\tau_{min} = 0.75$ and $\tau_{max} = 1.5$ being the minimum and maximum stretch factor.
To allow for $\tau>1.0$, mel-spectrograms are sampled with a temporal context of $4.5$\,s and truncated to $3$\,s after the interpolation.
Both $X[u,m]$ and $X_{TS}[u,m]$ are then processed by MULE to arrive at two vector representations 
$z=g(X[u,m])$ and $z'=g(X_{TS}[u,m])$.

Finally, the goal is to learn an embedding translation function $f$, parameterized by $\boldsymbol{\theta}$, that takes the original embedding $z$ and the stretch factor $\tau$ as input and predicts an estimation of $z'$, i.e.,
\begin{equation}
    \hat{z}' = f(z, \tau; \boldsymbol{\theta}).
\end{equation}

\begin{figure}[t!]
 \centerline{
 \includegraphics[width=0.75\columnwidth]{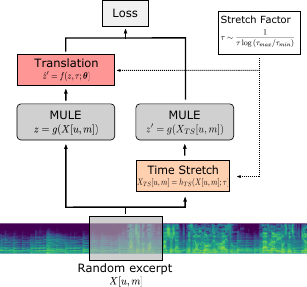}}
 \caption{Outline of the training setup. Given a mel-spectrogram excerpt, the task of the translation function is to predicted the translated embedding of a time-stretched version of said excerpt.}
 \label{fig:training_sketch}
\end{figure}

Learning $\boldsymbol\theta$ is entirely self-supervised without any need for labeled information by randomly constructing training triples ($z$, $\tau$, $z'$) and optimizing the sum of cosine similarity and mean-squared-error between the translated embedding $z'$ and the predicted translation $\hat{z}'$. 
An overview of the entire process is depicted in Figure~\ref{fig:training_sketch}. In our experiments, we use a translation function that was trained on an (unlabeled) set of $1.7M$ tracks, with two layers of size $2048$, a batch size of $256$, and the Adam optimizer \cite{kingma14adam}. The initial learning rate of $0.001$ is annealed to $0$ over the course of $200$k steps using a cosine learning rate scheduler with warm-up over $2000$ steps \cite{LoshchilovH17_SGDR_ICLR}.

We identify three specific efficiency advantages over other approaches that try to manipulate the embedding space itself, e.g., \cite{lee2020disentangled}, as it operates on a pre-existing embedding space. Firstly, training of the tempo translation network itself is much more efficient than the training of the embedding network because there is no need for backward propagation through this larger network as its weights are frozen. Secondly, nearest neighbor retrieval of tracks with a specific tempo, or along a contour of tempi only requires the relatively simple translation network to operate directly on embeddings without any reconstruction of those embeddings via the embedding model,\footnote{Our measurements indicate the translation model is 40$\times$ faster on a 32-core CPU (56ms per 128-batch) and 12$\times$ faster on an NVIDIA P100 GPU (40ms per 128-batch), than the embedding model. This brings the translation latency down to levels comparable to approximate nearest neighbours retrieval (i.e., $\approx 38$ms per 128 items).} as would be required if the translated embedding were to be re-constructed via time / frequency domain modification of the audio signal, each time. Finally, as a data augmentation strategy for training downstream tempo labelling probes, the entire embedding training dataset may be precomputed only once. During training only the lower complexity translation network is needed to augment embeddings, rather than modifying the audio features themselves and applying embedding model inference as part of the training process.

\section{Experiments \& Results}

In order to evaluate the efficacy of the embedding tempo translation model, we consider three potential downstream use cases: 
i) Nearest neighbor retrieval of embeddings with specific tempo; ii) nearest neighbor retrieval of tempo contours to retrieve neighbors impartial to tempo; and iii) data augmentation for training of downstream tempo probes.

\subsection{Nearest Neighbor Retrieval of Specific Tempo}
\label{subsec:nnr_tempo}

In some applications (e.g., in DJ software, music production software, or in some cases, playlist creation) a user may have a reference track that describes many of the qualities they are interested in, but at the ``wrong'' tempo. In these cases, the tempo translation network can be useful as an efficient way to modify the tempo encoded within an embedding, while maintaining its other properties. The modified embedding may be then used in an approximate nearest neighbor algorithm to 
retrieve audio of interest.
To demonstrate this, we consider two metrics, first, the \emph{Accuracy 1} score \cite{gouyon06taslp} between the tempo of a translated query embedding and its $k$-nearest neighbor's tempi, using $k=5$.    
In the case of embedding translation, we consider the query embedding to have the translated tempo (via. $T'=\tau T$), rather than that of the original audio. In this way, if the tempo translation of the embedding is successful, we should expect many of its K-nearest neighbors to have tempi close to the translated tempo, and hence observe high \textit{Accuracy 1}.
Secondly, to demonstrate the consistency of properties other than tempo in the translated embedding we consider the precision of its K-nearest neighbor's tags relative to its own, averaged over all labels of the query embedding. In both cases these metrics are computed over the neighborhood of all embeddings in the dataset, and averaged.

We compute these metrics for embedding translation factors from $0.5$ to $2.0$, over the Gtzan dataset \cite{tzanetakis2002musical, marchand2015swing} (for tempo) and Magnatagtune (MTT) \cite{law2009evaluation}  (for tags) datasets. As a baseline we consider the same metrics for embeddings reconstructed from audio that has been modified via a time stretching algorithm.\footnote{\url{https://sox.sourceforge.net/sox.html}}  While these embeddings are constructed from audio that is perceptually very similar to the source audio, they are much more computationally expensive to create. 
We also include the \emph{Accuracy 1} score of untranslated embeddings with translated tempi labels, i.e., the embeddings stay the same across the different translation factors and we only assume the tempo label to be changed.
This should demonstrate the change in tempo away from the desired translated value $T'$ if retrieval is attempted without any tempo translation, thereby measuring how many tracks with a specific translated tempo were within the initial neighborhood.

\begin{figure}[t!]
 \centerline{
 \includegraphics[width=1.\columnwidth]{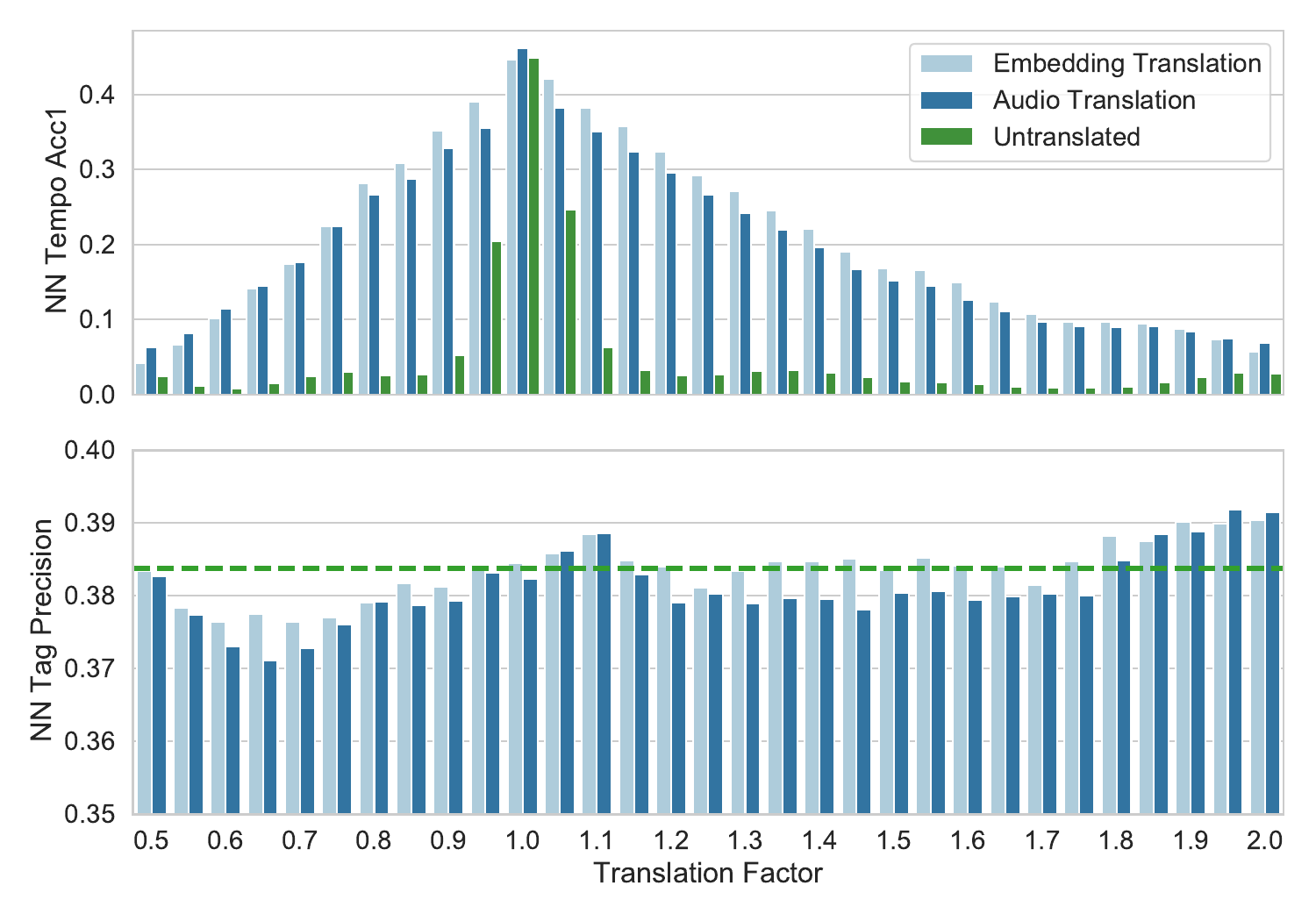}}
 \caption{Alignment between tempo and tags of the source embeddings and their $5$ nearest neighbors (NN) across different factors for embedding translation, audio translation (based on Sox time-stretching), and untranslated embeddings. Tempo alignment is
 reported on the Gtzan dataset and 
 is measured by \emph{Accuracy 1} (\emph{Acc1}) \cite{gouyon06taslp} between the translated source tempo and nearest neighbor tempi. For tag alignment, we report tag precision of neighbors on the test partition of the MTT dataset.
 }
 \label{fig:translation_verification}
\end{figure}

The results for this experiment are shown in Figure~\ref{fig:translation_verification}. There we can see that the translation network performs very well in achieving translated embeddings that are in neighborhoods with tempi similar to the desired / translated tempo, whilst maintaining the similarity in the neighbors tags. In both cases, we see that the tempo translation maintains a similar agreement with tags and tempo as the version where the audio is modified. As translation factors diverge in either direction, we see that the alignment of the embedding's tempo (either via translation, or audio time-stretching) with neighbors decay. At extreme tempo modifications we note that many tracks may reach areas that are unobserved in the training data, i.e., certain genres, moods and qualities of music at particular tempi may not exist.

\subsection{Nearest Neighbor Retrieval Impartial to Tempo}
\label{subsec:nnr_impartial}

The tempo translation network may not only be used to query embedding spaces at specific tempi, but across all tempi. In this way, we can find the nearest neighbors of an embedding contour that spans a range of tempi for a particular audio query. To achieve this by modifying the audio itself would be very computationally expensive, as it requires multiple STFT computations for time-stretching as well as the generation of the corresponding MULE embeddings. However with our proposed translation network, an embedding may be translated to several tempo values directly. 
By obtaining the neighbors that are close to the path of an embedding translated across a range of tempi, we hypothesise that we may find better agreement with the query track with respect to properties that are largely uncorrelated with tempo. To this end, we consider a sampled tempo contour in the embedding space.
\begin{equation}
z'[c] = f(z,1+c\delta;\boldsymbol{\theta})
\end{equation}
for indices $c \in [-C, C]$ where $C=10$ and at tempo increments $\delta=0.05$. For each $c$ we find the $k$-nearest neighbors, then out of those $(2C+1)k$ neighbors we further find the $k$ nearest neighbors that have the minimum distance to any point on the contour.

To evaluate, we consider the tag recall (as in \cite{lee2020metric}) across the test partition of the Million Song dataset (MSD) \cite{bertinMahieux2011msd}, following the widely used split with the 50 most common labels.\footnote{\url{https://github.com/jongpillee/music_dataset_split}} This tag recall is the percentage of embeddings with $k$-neighbors that have at least one tag in common with that embedding. We also consider for each of the labels of the query embedding, the precision of the labels of the retrieved $k$-nearest neighbors. This is averaged across all labels for each embedding and averaged across all embeddings in the dataset. We compare these metrics between the aforementioned nearest neighbors of each embedding's tempo contour. 

We consider three baselines. First, the nearest neighbors of the untranslated embedding. Secondly, the nearest neighbors of Gaussian clusters of $2C+1$ embeddings, using the same nearest neighbor method as for tempo contours. Clusters are created by additively augmenting each embedding with a Gaussian noise vector, $\hat{z} = z + \mathbf{g}_\sigma$, $2C+1$ times at standard deviations of $\sigma=0.1$ and $\sigma=0.5$. Thirdly, we consider the nearest neighbors of contours formed by sampling $2C+1$ points, linearly interpolated at equal intervals between $z'[-c_{max}]$ and $z'[c_{max}]$ where $c_{max}=C$. This third baseline indicates the importance of a non-linear translation function, $f(z,\tau;\boldsymbol{\theta})$. We could not find any significant improvement in this baseline when narrowing the tempo range, i.e., for $c_{max}<C$.

In Table~\ref{table:contourretrieval} we see that the MULE model alone preforms well in producing neighbors that have similar tags to the query, however, by considering a wider range of a track's tempi via its embedding tempo contour created by the tempo translation network, we are able to retrieve tracks that are, on average, in greater agreement with the query track in terms of their labels. This improvement does not hold for retrieval of nearest neighbors that are closest to any of a number of random permutation of embeddings via Gaussian noise, suggesting that it is not the regularization of random embedding augmentations that provides this increase in performance, but the successful navigation of the tempo translation network through the embedding space to regions that agree with the query track's other properties. We also see that while linear interpolation in the embedding domain helps, a non-linear translation function further improves results.

\begin{table}[t!]
  \centering
  \footnotesize
  \begin{tabular}{@{}cccccccccccc@{}}
    \multirow{2}{*}{\textbf{}} & \multicolumn{3}{c}{Precision} & \multicolumn{4}{c}{Tag Retrieval} \\
     & $k$=2 & $k$=4 & $k$=8 & $k$=1 & $k$=2 & $k$=4 & $k$=8 \\
    \toprule[1.1pt]
    MULE & 44.1 & 40.1 & 36.3 & 47.6 & 59.7 & 71.0 & 80.7  \\
    \midrule
    Tempo Contour & \textbf{49.4} & \textbf{45.3} & \textbf{40.8} & \textbf{52.8} & \textbf{65.6} & \textbf{77.2} & \textbf{86.0}  \\
    Gaussian \scalebox{0.9}{$\sigma=0.1$}  & 44.0 & 40.1 & 36.3 & 47.6 & 59.6 & 70.9 & 80.7 \\
    Gaussian \scalebox{0.9}{$\sigma=0.5$} & 42.5 & 38.6 & 34.9 & 45.4 & 57.7 & 69.4 & 79.6 \\
    Linear Interp. & 44.3 & 40.4 & 35.5 & 44.8 & 61.7 & 75.0 & 84.2 \\
    \midrule[1.1pt]
    \cite{lee2020metric}& - & - & - & 45.0 & 58.5 & 71.0 & 80.9 \\
    \bottomrule[1.1pt]
    \end{tabular}  
    \caption{Nearest neighbor content based retrieval performance. Metric definitions are equivalent for $k=1$.}
   \label{table:contourretrieval}
\end{table}

\subsection{Data augmentation for downstream tempo labelling}
\label{subsec:data_aug}
Finally, we use the translation function as an efficient data augmentation strategy to learn tempo predictors. A common technique for data augmentation in this domain is time stretching \cite{quinton22ismir, schreiber18ismir, boeck20ismir}, which is often done directly on the audio waveform or on spectrograms. However, both approaches can be computationally expensive depending on the complexity of the model used to process the audio.
As we are using a large network to create the embeddings used for the downstream classification, we want to avoid costly recomputations of these embeddings which would be necessary in case the underlying audio/spectrogram changes. Given the translation function, we have a direct way to change the tempo represented in the embedding, i.e., during training of a tempo classification model based on embeddings $z$ we randomly sample stretch factors $\tau$, compute the translated embeddings $z'$ and compute new tempo labels as $T'=\tau T$.

Using MULE embeddings as input, we train a multi-layer perceptron (MLP) to predict tempo, and frame the learning as a $271$ class classification task across the range of $30$ to $300$ beats per minute (BPM). 
 Specifically, the MLP consists of a single layer with $512$ neurons and a dropout rate of $0.75$. We train on batches of $256$ samples using Adam to optimize categorical crossentropy over $20$k steps. The learning rate is annealed from $0.0001$ to $0$ following a cosine learning rate scheduler with warm-up. MLPs are trained on the same collection of datasets as in \cite{boeck20ismir} with the addition of the Harmonix dataset \cite{nieto19ismir}. For evaluation, we hold out Gtzan \cite{tzanetakis2002musical, marchand2015swing}, Giantsteps-tempo \cite{schreiber2018crowdsourced, knees2015two}, and ACM-Mirum \cite{peeters2012perceptual, percival2014streamlined} as individual test sets and report \emph{Accuracy 1} and \emph{Accuracy 2} scores\cite{gouyon06taslp}. 
 
Table \ref{table:augmentation_results} summarizes our results and compares the translation augmentation to a variety of other strategies. In particular, we consider time-stretching of the mel-spectrogram (Mel-Augmentation) to see if the the translation augmentation can achieve a similar performance. Furthermore we compare against changes to the input embeddings that do not affect the label by applying dropout (with a drop rate, $p=0.25$) and additive Gaussian noise (with $\sigma=0.05$). 
As baseline approaches, we consider an end-to-end trained convolutional network \cite{schreiber18ismir}, a self-supervised tempo network \cite{quinton22ismir}, and a state-of-the-art bespoke model trained to jointly predict tempo, beat and downbeat in a multi-task setup \cite{boeck20ismir}.

We observe that both the translation and the mel-spectrogram augmentation boost the performance of the non-augmented MULE model. Given that both augmentations yield a similar performance we conclude that the direct augmentation of the embedding is a viable and efficient alternative to the classic augmentation strategy. By contrast, simply performing non-informed augmentation in the form of Gaussian noise and input dropout does not improve the results.
Finally, we note that our embedding-based approaches show a strong performance compared to state-of-the-art approaches, without being specifically designed for tempo estimation, further demonstrating the merit of powerful generic audio embeddings.

\begin{table}[t!]
\centering
\footnotesize
  \begin{tabular}{@{}ccccccccccc@{}}
    \multirow{2}{*}{\textbf{}} & \multicolumn{2}{c}{Gtzan} & \multicolumn{2}{c}{ACM-Mirum} & \multicolumn{2}{c}{Giantsteps} \\
  & Acc1        & Acc2        & Acc1          & Acc2          & Acc1           & Acc2          \\
\toprule[1.1pt]
  MULE         &   74.1 & 90.5 & 81.2 & 95.8 & 85.5 & \textbf{98.2} \\ \midrule
  Mel-Augmentation      &   77.7 & 91.6 & 82.1 & 96.2 & 90.3 & \textbf{98.2} \\
  Translation           &   77.7 & 92.1 & 83.6 & 95.7 & \textbf{90.7} & \textbf{98.2} \\
  Dropout \scalebox{0.9}{$p=0.25$}    &   74.3 & 90.7 & 81.2 & 95.6 & 83.2 & 98.0 \\
  Gaussian \scalebox{0.9}{$\sigma=0.05$}      &   74.4 & 90.5 & 81.5 & 95.8 & 84.4 & \textbf{98.2} \\
  Dropout + Gaussian       &   75.3 & 90.7 & 81.4 & 95.7 & 82.9 & 98.0 \\
         \midrule[1.1pt]       
    \cite{schreiber18ismir} & 76.9 & 92.6 & 78.1 & 97.6 & 82.1 & 97.1 \\
    \cite{quinton22ismir} \scalebox{0.9}{($r_f=0.2$)} & 74.1 & 91.9 & 74.7 & 96.5 & 47.0 & 88.6 \\
    \cite{boeck20ismir}& \textbf{83.0} & \textbf{95.0} & \textbf{84.1} & \textbf{99.0} & 87.0 & 96.5 \\
     \bottomrule[1.1pt]
\end{tabular}
     \caption{Tempo prediction performance as measured by \emph{Accuracy 1} (\emph{Acc1}) and \emph{Accuracy 2} (\emph{Acc2}) of the embedding-based approaches and different augmentation strategies compared to three baselines.}
\label{table:augmentation_results}
\end{table}

\section{Conclusions}
In this work, we present a first study on the direct manipulation of a property in the embedding space, using musical tempo as an example. We showed how learning a tempo translation function effectively enables and improves nearest neighbor retrieval of tracks with similar musical characteristics but different tempi. Furthermore, we are able to retrieve nearest neighbors that are impartial to tempo by obtaining neighbors along the embedding contour.
Finally, we use this translation function in a downstream tempo estimation task to efficiently augment the training data, obtaining strong performance compared to state-of-the-art baselines.

While tempo was chosen as an exemplary musical property to be manipulated in the embedding space, future work could investigate the manipulation of other characteristics relevant to music recommendation such as instrumentation, mood or genre.

\bibliographystyle{IEEEbib2} 
\bibliography{article}

\end{document}